# Exact Axisymmetric Solutions of the 2-D Lane-Emden Equations with Rotation

Dimitris M. Christodoulou[1], Demosthenes Kazanas[2]

[1]Department of Mathematical Sciences and Lowell Center for Space Science and Technology, University of Massachusetts Lowell, Lowell, USA
[2]NASA Goddard Space Flight Center, Laboratory for High-Energy Astrophysics, Greenbelt, USA
Email: dimitris_christodoulou@uml.edu, demos.kazanas@nasa.gov

## Abstract

We have derived exact axisymmetric solutions of the two-dimensional Lane-Emden equations with rotation. These solutions are intrinsically favored by the differential equations regardless of any adopted boundary conditions and the physical solutions of the Cauchy problem are bound to oscillate about and remain close to these intrinsic solutions. The isothermal solutions are described by power-law density profiles in the radial direction, whereas the polytropic solutions are described by radial density profiles that are powers of the zeroth-order Bessel function of the first kind. Both families of solutions decay exponentially in the vertical direction and both result in increasing or nearly flat radial rotation curves. The results are applicable to gaseous spiral-galaxy disks that exhibit flat rotation curves and to the early stages of protoplanetary disk formation before the central star is formed.

## 1. Introduction

We use a new method to solve analytically the axisymmetric Lane-Emden equations [1] [2] with rotation in two dimensions. The method is an extension of the one-dimensional algorithm that we applied to ordinary second-order differential equations of mathematical physics [3] [4] [5] and produces separable equations in two dimensions [6]. The solutions are intrinsically favored by the differential equations themselves and dictate that the physical solutions of the Cauchy problem should oscillate about and remain close to these preferred solutions [4] [5].

The two-dimensional analytic solutions show that both the densities and the rotation speeds decay exponentially with height from the symmetry plane $z = 0$ while the radial rotation curves are increasing or nearly flat at all heights. Thus, the Newtonian rotation profiles so derived are similar to the "flat" rotation curves observed in gaseous spiral galaxies [7] without the need of invoking dark matter [8]-[22] or the various modifications of the Newtonian dynamics [23]-[30].

In what follows, we derive the exact solutions of the 2-D Lane-Emden equations with rotation in the isothermal case (Section 2) and in the general polytropic case (Section 3), and we discuss the astrophysical implications of our results (Section 4).

## 2. Isothermal Self-Gravitating Newtonian Gaseous Disks

We use the scaling constants $R_o$ and $\rho_o$ to normalize the cylindrical coordinates ($R$, $Z$) and the density profiles $\rho(R,Z)$, respectively. We thus define the dimensionless radius $x \equiv R/R_o$, height $z \equiv Z/R_o$, and density $\tau(x,z) \equiv \rho(R,Z)/\rho_o$. Velocities $V(R,Z)$ are also normalized consistently by the constant $V_o = R_o\sqrt{4\pi G \rho_o}$, where $G$ is the Newtonian gravitational constant, in which case we define the dimensionless rotation velocity $v(x,z) \equiv V(R,Z)/V_o$. The same scaling also applies to the sound speed $C_o$ of the gas which in this section is a constant, $i.e.$, the dimensionless sound speed is $c_o \equiv C_o/V_o$.

The 2-D axisymmetric isothermal Lane-Emden equation with rotation [1] [2] [5] can then be written in dimensionless form as

$$c_o^2\left[\left(\nabla_x^2 + \nabla_z^2\right)\ln\tau\right] + \tau = \frac{1}{x}\frac{\partial v^2}{\partial x}, \tag{1}$$

where $\tau$ and $v$ are functions of $x$ and $z$, and

$$\nabla_x^2 \equiv \frac{1}{x}\frac{\partial}{\partial x}x\frac{\partial}{\partial x}, \tag{2}$$

and

$$\nabla_z^2 \equiv \frac{\partial^2}{\partial z^2}. \tag{3}$$

This equation describes the axisymmetric equilibrium of a rotating, self-gravitating, gaseous disk in which the gas obeys the isothermal equation of state $p(x,z) = c_o^2\,\tau(x,z)$, where $p$ is the dimensionless pressure of the gas.

If we equate the last two terms of Equation (1), viz.

$$\tau(x,z) = \frac{1}{x}\frac{\partial v^2(x,z)}{\partial x}, \tag{4}$$

then this is an intrinsic solution [4] [5] provided that the rest of the equation also vanishes:

$$\left(\nabla_x^2 + \nabla_z^2\right)\ln\tau(x,z) = 0. \tag{5}$$

Equations ((4) and (5)) form a system in which $v(x,z)$ is determined from $\tau(x,z)$ which, in turn, is determined by solving the Laplace Equation (5).

We now introduce the following scaling relations in the $z$-direction:

$$\begin{cases} \tau(x,z) = y(x)f(z) \\ v(x,z) = s(x)\sqrt{f(z)} \end{cases}, \qquad (6)$$

where the three new functions $y(x)$, $s(x)$, and $f(z)$ are to be determined self-consistently from Equations (4) and (5). Combining Equations (4) and (6), we find that at every height $z$, the radial variation of the rotation velocity $s(x)$ is determined from an integral of the radial density function $y(x)$ using the equation

$$y(x) = \frac{1}{x}\frac{\mathrm{d}s^2(x)}{\mathrm{d}x}. \qquad (7)$$

This is precisely the equation that was solved by [5] on the symmetry plane $z = 0$ of the disk. We proceed now to solve for the $z$-dependence in Equations ((1) to (6)). Substituting the first of Equations (6) into Equation (5), we find that

$$\nabla_x^2 \ln y(x) + \nabla_z^2 \ln f(z) = 0. \qquad (8)$$

The two terms are independent, thus they must be constant and the constants should combine to produce zero. We can then write

$$\nabla_x^2 \ln y(x) = -\nabla_z^2 \ln f(z) = \mu^2, \qquad (9)$$

where the separation constant $\mu^2$ is taken to be positive (or zero) to ensure that $f(z)$ is a decreasing function of $|z|$. Integrating these two equations separately, we find for $f(0) = 1$ that

$$f(z) = \exp\left(-\frac{1}{2}\mu^2 z^2 + c|z|\right) \quad (c < 0), \qquad (10)$$

where the integration constant $c$ is taken to be negative to guarantee monotonically decreasing density profiles away from the symmetry plane $z = 0$; and that

$$y(x) = Ax^{k-1}\exp\left(+\frac{1}{4}\mu^2 x^2\right) \quad (A > 0, k < 1), \qquad (11)$$

where $A, k-1$ are the integration constants. Because of the exponential term, the function $y(x)$ is asymptotically increasing with $x$, and this leads to radially increasing density profiles. These solutions are unphysical and we are forced to choose

$$\mu = 0, \qquad (12)$$

in which case the solutions become

$$f(z) = \exp(-|cz|), \qquad (13)$$

and

$$y(x) = Ax^{k-1} \quad (A > 0, k < 1). \qquad (14)$$

Equation (14) was derived by [5] for $z = 0$, whereas in this treatment, equation (13) describes the solutions for heights away from the equatorial plane. The solutions in the entire $(x, z)$ plane take the form

$$\begin{cases} \tau(x,z) = Ax^{k-1}\exp(-|cz|) \\ v(x,z) = s(x)\exp\left(-\frac{1}{2}|cz|\right) \end{cases} \quad (A > 0, k < 1), \qquad (15)$$

where $s(x)$ is determined from Equations ((7) and (14)) as was done in [5], where it was shown that all rotation curves $s(x)$ are slowly increasing or flat with radius $x$. This feature of the rotation profiles remains valid away from $z = 0$ despite the exponential decay of the density $\tau(x,z)$ with $|z|$.

Figure 1 shows the density profile $\log \tau(x,z)$ of the isothermal Lane-Emden equation for $k = -1$ and $c = -10$ (Equation (15)). The choice $c = -10$ causes a rapid decline of the densities with height and gives the model the appearance of a disk-like structure that is centrally condensed because of the steep $1/x^2$ radial dependence of the density. The rotation curve of this solution was shown in [5] where $s^2(x)$ increases logarithmically with radius for $x \geq 1$.

## 3. Polytropic Self-Gravitating Newtonian Gaseous Disks

The 2-D axisymmetric polytropic Lane-Emden equation with rotation [1] [2] [5] can be written in dimensionless form as

$$nc_o^2\left[\left(\nabla_x^2 + \nabla_z^2\right)\tau^{1/n}\right] + \tau = \frac{1}{x}\frac{\partial v^2}{\partial x}, \qquad (16)$$

where $n > 0$ is the polytropic index and the dimensionless constant sound speed $c_o$ was defined for $\rho = \rho_o$. (In general, the square of the sound speed $c^2(x,z) \equiv dp/d\tau$ varies as $\tau^{1/n}$ across the medium, where, again, $p(x,z)$ is the dimensionless pressure of the gas.) This equation describes the axisymmetric equilibrium of a rotating, self-gravitating, gaseous disk in which the gas obeys a polytropic equation of state of the form $p \propto \tau^{1+1/n}$.

We repeat the procedure outlined in Section 2 in order to obtain the intrinsic solution of Equation (16): If we equate the last two terms of Equation (16), viz.

$$\tau(x,z) = \frac{1}{x}\frac{\partial v^2(x,z)}{\partial x}, \qquad (17)$$

then this is an intrinsic solution [4] [5] provided that the rest of the equation also vanishes:

$$\left(\nabla_x^2 + \nabla_z^2\right)\tau^{1/n}(x,z) = 0. \qquad (18)$$

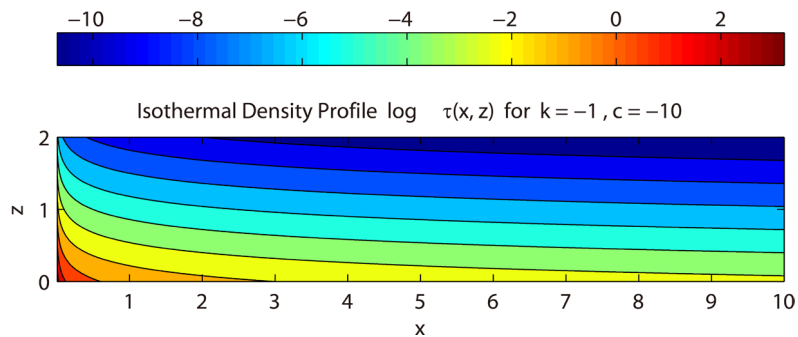

**Figure 1.** Density solution $\log \tau(x, z \geq 0)$ of the isothermal Lane-Emden equation for $k = -1$ and $c = -10$ (Equation (15)). Ten contours are plotted with the colors in the red part of the spectrum representing higher densities. The aspect ratio of the plot is set to $x:z = 5:1$.

Equations (17) and (18) form a system in which $v(x,z)$ is determined from $\tau(x,z)$ which, in turn, is determined by solving the Laplace Equation (18).

We now introduce the scaling relations (6) in the $z$-direction, where the three functions $y(x)$, $s(x)$, and $f(z)$ are to be determined self-consistently from Equations ((17) and (18)). Combining Equations ((17) and (6)), we find that at every height $z$, the radial variation of the rotation velocity $s(x)$ is determined from an integral of the radial density function $y(x)$ using the equation

$$y(x) = \frac{1}{x}\frac{\mathrm{d}s^2(x)}{\mathrm{d}x}, \tag{19}$$

as was also found in the isothermal case of Section 2.

Substituting the first of Equations (6) into Equation (18) and diving all terms by $(yf)^{1/n}$, we find that

$$y^{-1/n}\nabla_x^2 y^{1/n}(x) + f^{-1/n}\nabla_z^2 f^{1/n}(z) = 0. \tag{20}$$

The two terms are independent, thus they must be constant and the constants should combine to produce zero. We can then write

$$y^{-1/n}\nabla_x^2 y^{1/n}(x) = -f^{-1/n}\nabla_z^2 f^{1/n}(z) = -\mu^2, \tag{21}$$

where the separation constant $-\mu^2$ is taken to be negative to ensure that $y(x)$ is not a monotonically increasing function of $x$ and is not singular at $x=0$. Such solutions (the zeroth-order modified Bessel functions of the first kind $I_0(\mu x)$ described in [31]) are obtained for positive separation constants, whereas the case $\mu = 0$ produces the singular solutions found in [5]. Integrating the two Equations (21) separately, we find for $f(0) = 1$ that

$$f(z) = \exp(-n\mu|z|) \quad (\mu > 0), \tag{22}$$

where the particular solution with the minus sign was chosen so that $f(z)$ decreases with $|z|$; and that

$$y(x) = A[J_0(\mu x)]^n \quad (A > 0), \tag{23}$$

where $A$ is the integration constant and the Bessel function of the first kind was chosen because it does not diverge at $x=0$. These solutions are monotonically decreasing with $x$ and terminate at the first zero of the Bessel function $J_0(\mu x)$, viz. [12]

$$x_1 = \frac{2.4048}{\mu}, \tag{24}$$

except in cases of even polytropic indices $n$ in which they produce rings touching one another at consecutive zeroes of $J_0(\mu x)$.

The solutions in the entire ($x,z$) plane take the form

$$\begin{cases} \tau(x,z) = A[J_0(\mu x)]^n \exp(-n\mu|z|) \\ v(x,z) = s(x)\exp\left(-\frac{1}{2}n\mu|z|\right) \end{cases} (A, n, \mu > 0), \tag{25}$$

where $s(x)$ is determined from Equations (19) and (23), viz.

$$s(x) = \sqrt{A \int_0^x x \left[ J_0(\mu x) \right]^n \, \mathrm{d}x}, \tag{26}$$

where $s(0) = 0$. The integral in Equation (26) can be written analytically in terms of the Bessel functions of the first kind $J_0(\mu x)$ and $J_1(\mu x)$ for some integer values of $n$ as follows [31] [32]:

$$\int_0^x x \left[ J_0(\mu x) \right]^n \mathrm{d}x$$

$$= \begin{cases} \mu^{-1} x J_1(\mu x) \ (x \leq x_1), & \text{for } n = 1, \\ \dfrac{1}{2} x^2 \left[ J_0^2(\mu x) + J_1^2(\mu x) \right], & \text{for } n = 2, \\ \mu^{-1} \left[ x J_0^2(\mu x) J_1(\mu x) + \dfrac{2x}{3} J_1^3(\mu x) + \dfrac{4}{3} \int J_1^3(\mu x) \mathrm{d}x \right] (x \leq x_1), & \text{for } n = 3, \end{cases} \tag{27}$$

where $x_1$ is given by Equation (24).

Figure 2 and Figure 3 show the density and rotation profiles, respectively, of the $n = 2$ Lane-Emden equation for $\mu = 0.1$ (Equation (25)). These profiles are representative of other solutions as well with $\mu < 0.1$, whereas for even values of $n$, larger values of $\mu$ produce differentially rotating ring-like structures in which the rings touch one another at the zeroes of the Bessel function $J_0(\mu x)$. An example of such a ring solution with $n = 2$ and $\mu = 1$ and its rotation profile are shown in **Figure 4** and **Figure 5**.

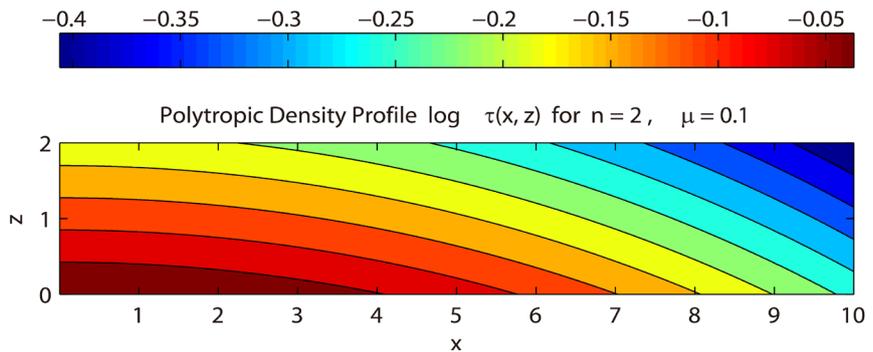

**Figure 2.** Density solution $\log \tau(x, z \geq 0)$ of the $n = 2$ polytropic Lane-Emden equation for $\mu = 0.1$ (Equation (25)). Eleven contours are plotted with the colors in the red part of the spectrum representing higher densities. The aspect ratio of the plot is set to $x:z = 5:1$.

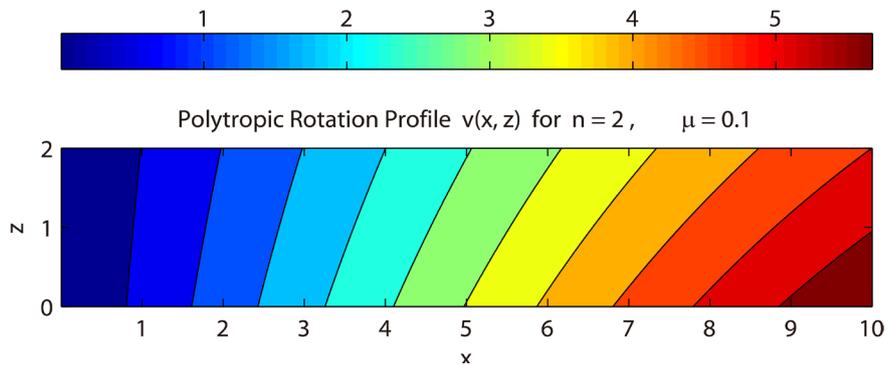

**Figure 3.** Rotation profile $v(x, z \geq 0)$ of the $n = 2$ polytropic Lane-Emden equation for $\mu = 0.1$ (Equation (25)). Eleven contours are plotted with the colors in the red part of the spectrum representing higher rotation speeds. The aspect ratio of the plot is set to $x:z = 5:1$.

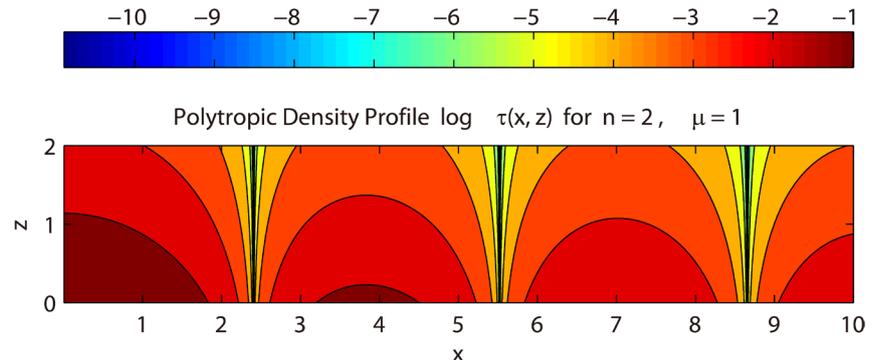

**Figure 4.** Ring-like density solution $\log \tau(x, z \geq 0)$ of the $n = 2$ polytropic Lane-Emden equation for $\mu = 1$ (Equation (25)). Ten contours are plotted with the colors in the red part of the spectrum representing higher densities. The aspect ratio of the plot is set to $x : z = 5 : 1$.

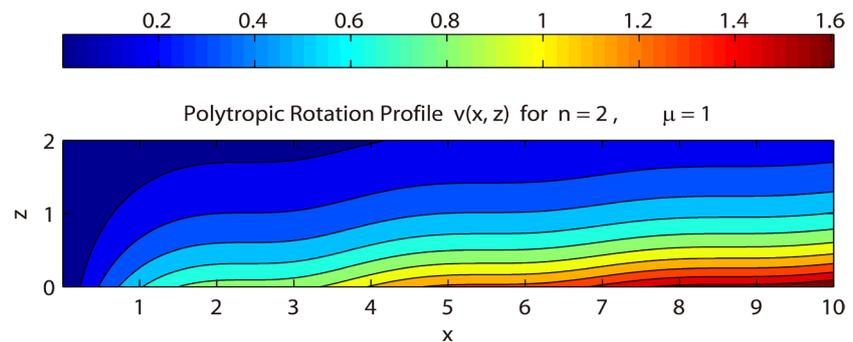

**Figure 5.** Rotation profile $v(x, z \geq 0)$ of the $n = 2$ polytropic Lane-Emden equation for $\mu = 1$ (Equation (25)). Eleven contours are plotted with the colors in the red part of the spectrum representing higher rotation speeds. The aspect ratio of the plot is set to $x : z = 5 : 1$.

In the inner region of the rotation profile of **Figure 3**, the contours are nearly vertical as expected from measurements of the rotation profile of the Milky Way [33]. In the outer region where the rotation speeds are larger, the contours are however tilted and the tilt becomes more pronounced for larger values of $\mu$. In this region, nearly vertical contours can, however, be produced for smaller values of $\mu < 0.1$, so it does not appear to be difficult to produce a Newtonian rotation profile such as that observed in our Galaxy for heights $|z| \leq 100$ pc [33]. An example of such a rotation profile with nearly vertical contours at all radii is shown in **Figure 6** for $n = 2$ and $\mu = 0.01$.

**Figure 7** shows the radial rotation curve $s(x)$ for $\mu = 0.1$ and $A = 1$, and for the cases $n = 1, 2,$ and $3$ (Equations ((26) and (27))). The $n = 1$ and $n = 3$ curves terminate at the first zero of the Bessel function $x_1$ (Equation (24)). These curves rise in the inner region and then they become asymptotically flat. The flat segments can be extended farther out in radius if values of $\mu < 0.1$ are used. In contrast, the $n = 2$ curve continues to increase with radius as it passes through a sequence of inflection points. Each jump in the profile represents the rotation curve inside the next outward ring.

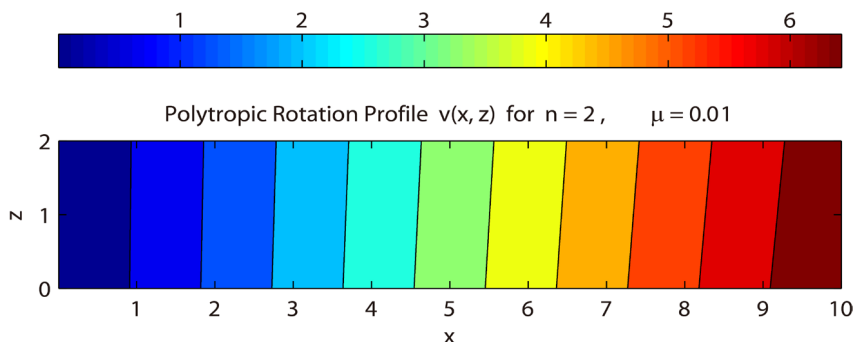

**Figure 6.** Rotation profile $v(x, z \geq 0)$ of the $n = 2$ polytropic Lane-Emden equation for $\mu = 0.01$ (Equation (25)). Eleven contours are plotted with the colors in the red part of the spectrum representing higher rotation speeds. The aspect ratio of the plot is set to $x:z = 5:1$. Compared to Figure 3, the smaller value of $\mu$ causes the contours here to become nearly vertical at all radii.

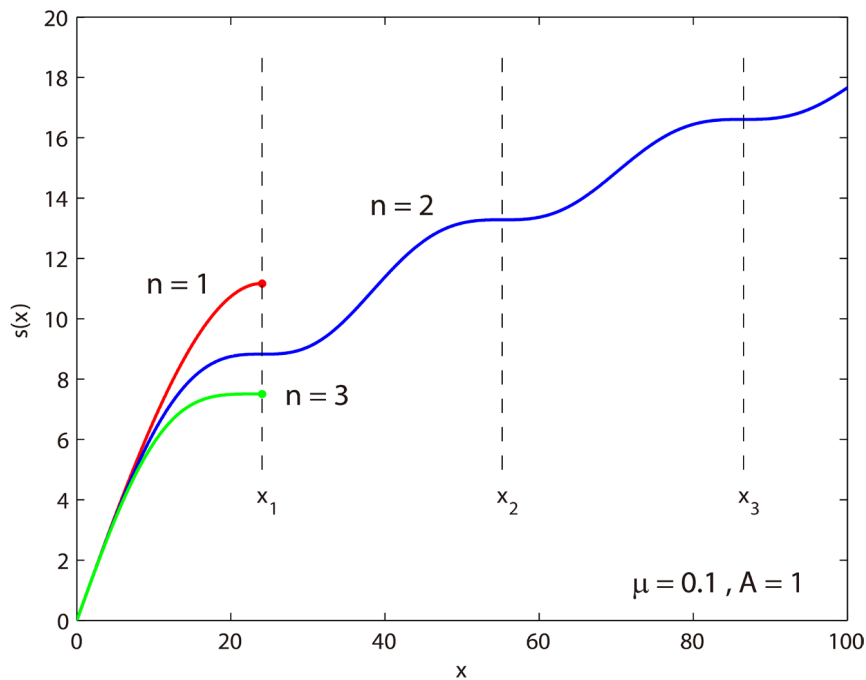

**Figure 7.** Polytropic rotation curve $s(x)$ as a function of radius $x$ for $n = 1, 2$, and 3, $\mu = 0.1$, and $A = 1$ (Equations ((26) and (27))). The zeroes of the Bessel function $J_0(\mu x)$ are shown by dashed lines. The first zero $x_1$ is given by Equation (24), $x_2 = 5.5201/\mu$, and $x_3 = 8.6537/\mu$. The $n = 1$ and $n = 3$ solutions terminate at the first zero $x_1$ where the density goes to zero.

## 4. Discussion

In this work, we derived exact axisymmetric solutions of the 2-D Lane-Emden equations with rotation. In the isothermal case, the solutions show a power-law dependence on the radius $x$ and an exponential decline with height $|z|$. In the general polytropic case with index $n > 0$, the radial solutions depend on powers of the zeroth-order Bessel function of the first kind $(J_0)^n$ and they also decline with $|z|$. Both families of solutions are intrinsic to the differential equations themselves; any solutions that obey physical boundary conditions will have to remain close to these solutions and they will be forced to oscillate about them if the boundary conditions will be different than the conditions that produced the intrinsic solutions [4] [5].

The gaseous disk equilibria produced in Sections 2 and 3 are Newtonian in nature and they all have rising or asymptotically flat radial rotation curves. Such rotation profiles are demanded by the analytic solutions for self-consistency. The resulting Newtonian models argue against the need to assume the existence of dark matter in spiral galaxies in order to produce the observed flat rotation curves and against the need to modify the Newtonian dynamics to achieve the same effect (references were given in Section 1). The same models may also prove useful in studies of the rotation curves in protoplanetary disks [34] [35] [36] at their very early stages of evolution and before the central star is formed.

It is important to note that the flat and rising rotation profiles were not prescribed as input to the Lane-Emden equations; instead, they were the result of the intrinsic equilibrium solutions. Similar types of rotation profiles have been previously found in models of Newtonian gaseous disks [37] [38] [39] [40] but they were dismissed because they were thought to be peculiar in nature. We now find that these models were giving us clues as to the true behavior of the gas in self-gravitating astrophysical disks.

The main obstacle in understanding the dynamical behavior of gas in spiral galaxies is an old argument [7] that relies solely on particle dynamics—that an orbiting particle at radius $r$ enclosing mass $M(r)$ must experience a rotation speed of $v = \sqrt{GM(r)/r}$ in equilibrium, thus only a mass $M(r) \propto r$ can produce a flat rotation curve. This argument is invalid for gaseous disks where the enthalpy of the gas controls the dynamics absolutely [5]. The intrinsic solutions derived here and in [5] show that the specific enthalpy of the gas $h(x,z) \equiv \int dp/\tau$ arranges the local density profile $\tau(x,z)$ in equilibrium and, subsequently, it is this density that sources and manipulates both the self-gravitational potential (via Poisson's equation) and the rotational potential (via Equation (17)). Thus, naive arguments that rely on massless particles reacting to a prescribed gravitational potential simply do not describe the dynamics of the gas which is entirely determined by the local distribution of the thermodynamical potential $h(x,z)$.

## Acknowledgements

We thank Joel Tohline for feedback and guidance over many years.

## References

[1] Lane, J.H. (1870) *Amer. J. Sci. Arts*, *Second Series*, **L**, 57.

[2] Emden, R. (1907) Gaskugeln, Leipzig, B.G. Teubner.

[3] Christodoulou, D.M., Graham-Eagle, J. and Katatbeh, Q.D. (2016) *Advances in Difference Equations*, **2016**, 48. https:/doi.org/10.1186/s13662-016-0774-x

[4] Christodoulou, D.M., Katatbeh, Q.D. and Graham-Eagle, J. (2016) *Journal of Inequalities and Applications*, **2016**, 147. https:/doi.org/10.1186/s13660-016-1086-0

[5] Christodoulou, D.M. and Kazanas, D. (2016) *Journal of Modern Physics*, **7**, 680-698. https:/doi.org/10.4236/jmp.2016.77067

[6] Jackson, J.D. (1962) Classical Electrodynamics. John Wiley & Sons, New York.

[7] Binney, J. and Tremaine, S. (1987) Galactic Dynamics. Princeton Univ. Press, Princeton.

[8] Freeman, K.C. (1970) *Astrophysical Journal*, **160**, 811. https:/doi.org/10.1086/150474

[9] Bosma, A. (1978) The Distribution and Kinematics of Neutral Hydrogen in Spiral Galaxies of Various Morphological Types. Ph.D. Thesis, University of Groningen, Groningen.

[10] Rubin, V.C., Ford Jr., W.K. and Thonnard, N. (1980) *Astrophysical Journal*, **238**, 471. https:/doi.org/10.1086/158003

[11] Bosma, A. (1981) *Astronomical Journal*, **86**, 1791-1846. https:/doi.org/10.1086/113062

[12] Bosma, A. (1981) *Astronomical Journal*, **86**, 1825-1846. https:/doi.org/10.1086/113063

[13] Rubin, V.C., Ford Jr., W.K., Thonnard, N. and Burstein, D. (1982) *Astrophysical Journal*, **261**, 439-456. https:/doi.org/10.1086/160355

[14] Van Albada, T.S. and Sancisi, R. (1986) *Philosophical Transactions of the Royal Society of London A*, **320**, 447-464. https:/doi.org/10.1098/rsta.1986.0128

[15] Begeman, K.G. (1987) HI Rotation Curves of Spiral Galaxies. PhD Thesis, University of Groningen, Groningen.

[16] Begeman, K.G. (1989) *Astronomy & Astrophysics*, **223**, 47-60.

[17] Persic, M. and Salucci, P. (1990) *Monthly Notices of the Royal Astronomical Society*, **245**, 577.

[18] Carignan, C., Charbonneau, P., Boulanger, F. and Viallefond, F. (1990) *Astronomy & Astrophysics*, **234**, 43-52.

[19] Broeils, A. (1992) Dark and Visible Matter in Spiral Galaxies. PhD Thesis, University of Groningen, Groningen.

[20] Persic, M. and Salucci, P. (1995) *Astrophysical Journal Supplement*, **99**, 501. https:/doi.org/10.1086/192195

[21] Persic, M., Salucci, P. and Stel, F. (1996) *Monthly Notices of the Royal Astronomical Society*, **281**, 27-47. https:/doi.org/10.1093/mnras/278.1.27

[22] Salucci, P. and Persic, M. (1997) Dark Halos around Galaxies. In: Persic, M. and Salucci, P., Eds., *Dark and Visible Matter in Galaxies*, ASP Conference Series 117, 1.

[23] Milgrom, M. (1983) *Astrophysical Journal*, **270**, 365-370. https:/doi.org/10.1086/161130

[24] Tohline, J.E. (1983) Stabilizing a Cold Disk with a 1/r Force Law. In: Athanassoula, E., Ed., *Internal Kinematics and Dynamics of Galaxies*, Springer, Berlin, 205-206. https:/doi.org/10.1007/978-94-009-7075-5_56

[25] Felten, J.E. (1984) *Astrophysical Journal*, **286**, 3-6. https:/doi.org/10.1086/162569

[26] Sanders, R.H. (1984) *Astronomy & Astrophysics*, **136**, L21-L23.

[27] Sanders, R.H. (1986) *Monthly Notices of the Royal Astronomical Society*, **223**, 539-555. https:/doi.org/10.1093/mnras/223.3.539

[28] Mannheim, P.D. and Kazanas, D. (1989) *Astrophysical Journal*, **342**, 635-638.

https:/doi.org/10.1086/167623

[29] Mannheim, P.D. and O'Brien, J.G. (2011) *Physical Review Letters*, **106**, Article ID: 121101. https:/doi.org/10.1103/PhysRevLett.106.121101

[30] Mannheim, P.D. and O'Brien, J.G. (2012) *Physical Review D*, **85**, Article ID: 124020.

[31] Abramowitz, M. and Stegun, I.A. (1972) Handbook of Mathematical Functions with Formulas, Graphs, and Mathematical Tables. Dover, New York.

[32] Rosenheinrich, W. (2016) Tables of Some Indefinite Integrals of Bessel Functions. http://www.eah-jena.de/rsh/Forschung/Stoer/besint.pdf

[33] Levine, E.S., Heiles, C. and Blitz, L. (2008) *Astrophysical Journal*, **679**, 1288-1298. https:/doi.org/10.1086/587444

[34] Williams, J.P. and Cieza, L.A. (2011) *Annual Review of Astronomy and Astrophysics*, **49**, 67-117. https:/doi.org/10.1146/annurev-astro-081710-102548

[35] Belloche, A. (2013) Observation of Rotation in Star Forming Regions: Clouds, Cores, Disks, and Jets. In: Hennebelle, P. and Charbonnel, C., Eds., *Angular Momentum Transport during Star Formation and Evolution*, EAS Pub. Ser. 62, 25.

[36] Tsitali, A.E., Belloche, A., Commerçon, B. and Menten, K.M. (2013) *Astronomy & Astrophysics*, **557**, Article Number: A98. https:/doi.org/10.1051/0004-6361/201321204

[37] Hayashi, C., Narita, S. and Miyama, S.M. (1982) *Progress of Theoretical Physics*, **68**, 1949-1966. https:/doi.org/10.1143/PTP.68.1949

[38] Narita, S., Kiguchi, M., Miyama, S.M. and Hayashi, C. (1990) *Monthly Notices of the Royal Astronomical Society*, **244**, 349-356.

[39] Schneider, M. and Schmitz, F. (1995) *Astronomy & Astrophysics*, **301**, 933-940.

[40] Marr, J.H. (2015) *Monthly Notices of the Royal Astronomical Society*, **448**, 3229-3241. https:/doi.org/10.1093/mnras/stv216

Scientific Research Publishing